# Wireless Sensor Network based Future of Telecom Applications

Authored By
Prof.(Dr.) Arun Dua
Head of CS/IT Department,
Ansal Institute of Technology,
Gurgaon, Haryana , India

## Abstract:

A system and method for enabling human beings to communicate by way of their monitored brain activity. The brain activity of an individual is monitored and transmitted to a remote location (e.g. by satellite). At the remote location, the monitored brain activity is compared with pre-recorded normalized brain activity curves, waveforms, or patterns to determine if a match or substantial match is found. If such a match is found, then the computer at the remote location determines that the individual was attempting to communicate the word, phrase, or thought corresponding to the matched stored normalized signal.

# Introduction:

Upanishad says that Human has Speech of Speech, Mind of Mind, and Hearing of Hearing. In other words, what we speak is part of what we intend to speak, what we hear and retained for usage is part of what all we heard and what our mind reflects, retain and processes for action is also part of what it intends out of many options.

In fact in current paradigm of Object Oriented terms there is multiple inheritance of observing(viewing), hearing and earlier learning's to mind and single inheritance as output in terms of speech and Further More Thoughts..

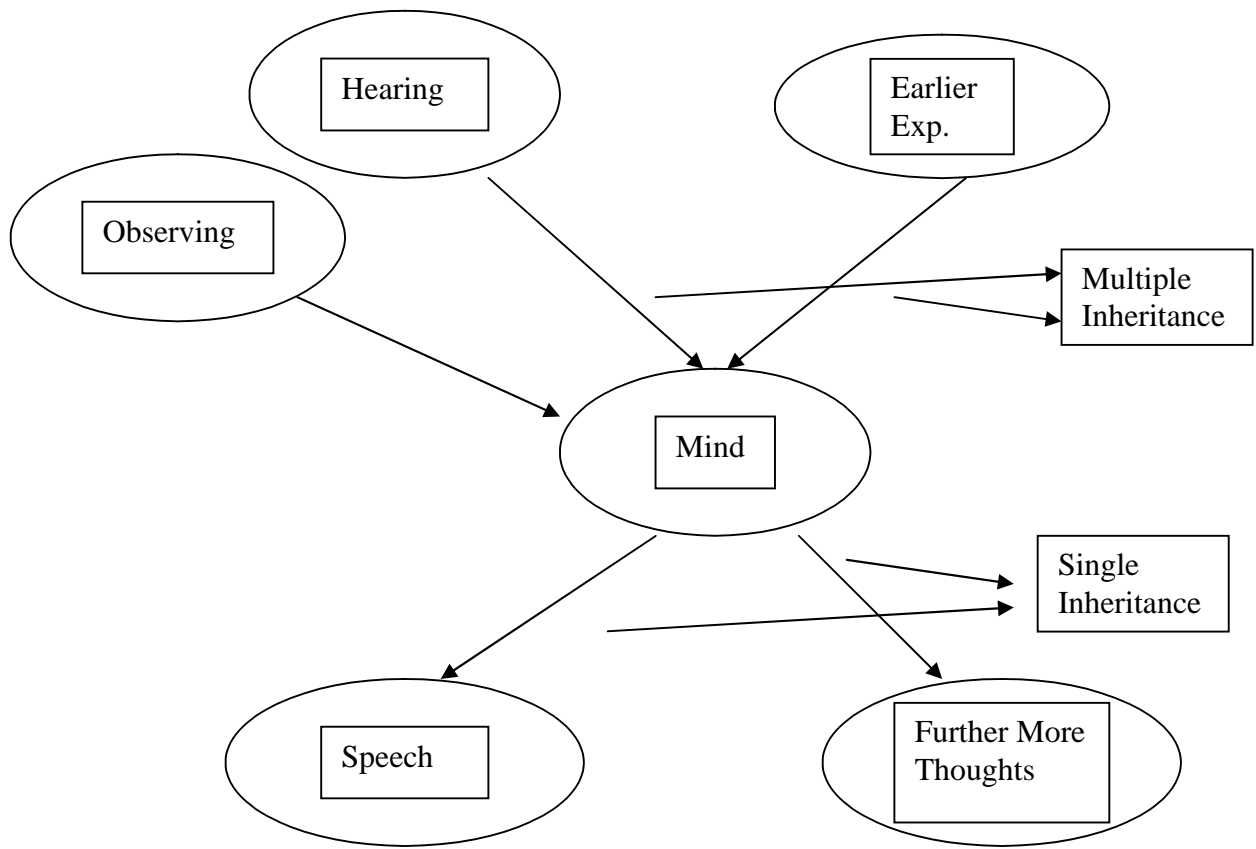

Human can have up to 4000 Hertz (Hz) as speech but can have listening up to 20000 Hz. Precisely due to this while we can listen loudspeaker outputs and get disturb/enjoy the generated output, we need microphone to speak and take care of attenuation of otherwise 5 times lesser speech frequency already.

While POTS (Plain Old Telephone Set) capture around 500 Hz and does further transmission through lines for present day communication. Less than 300Hz has been identified as minimum to make sound audible and traversing few feet's.

What about 0-300 Hz and what about sub zero Hz which are going in mind continuously but frequency being low are termed as running thought. If our POTS by putting repeaters can communicate why not repeaters for sub zero Hz capturing. Such repeaters would be wireless and WiFi and Bluetooth would come handy to capture. A Head Band or cap can have such WiFi and Bluetooth tools because our sub zero Hz generated as thought would require due to distance of few inches from brain to mouth.

Voila, our thoughts now can be captured as raw and we can generate speech of speech and hearing of hearing better by using computers rather using mind filters anymore. Mind being most precious and ageless would work for other new thoughts and functions like high utilization of CPU of our computers. Mind as a Human CPU is having the same fate of computer CPU that is most powerful in computer but least utilized.

## Description:

This future Wireless Sensor Network (WSN) based communication technology direction relates to a system and method for enabling human beings to communicate with one another by monitoring brain activity. In particular, this relates to such a system and method where brain activity of a particular individual is monitored and transmitted in a wireless manner (e.g. via satellite) from the location of the individual to a remote location so that the brain activity can be computer analyzed at the remote location thereby enabling the computer and/or individuals at the remote location to determine what the monitored individual was thinking or wishing to communicate.

In certain embodiments this future WSN based communication technology direction would relate to the analysis of brain waves or brain activity, and/or to the remote firing of select brain nodes in order to produce a predetermined effect on an individual.

## Summary:

Generally speaking, this future WSN Based communication technology direction fulfills the above described needs in the art by providing a method of communicating comprising the steps of:

- providing a first human being at a first location;

- providing a computer at a second location that is remote from the first location;

- providing a satellite;

- providing at least one sensor (preferably a plurality--e.g. tens, hundreds, or thousands, with each sensor monitoring the firing of one or more brain nodes or synapse type members) on the first human being;

- detecting brain activity of the first human being using at least one sensor, and transmitting the detected brain activity to the satellite as a signal including brain activity information;

- the satellite sending a signal including the brain activity information to the second location;

- a receiver at the second location receiving the signal from the satellite and forwarding the brain activity information in the signal to the computer;

- comparing the received brain activity information of the first human being with normalized or averaged brain activity information relating to the first human being from memory; and

- determining whether the first human being was attempting to communicate particular words, phrases or thoughts, based upon the comparing of the received brain activity information to the information from memory.

In certain terms, the WSN based future communication technology direction includes the following steps:

- asking the first human being a plurality of questions and recording brain activity of the first human being responsive to the plurality of questions in the process of developing said normalized or averaged brain activity information relating to the first human being stored in the memory.
- A database in a memory may include, for each of a plurality (e.g. one hundred or thousands) of individuals, a number of prerecorded files each corresponding to a particular thought, attempt to communicate a word, attempt to communicate a phrase or thought, or mental state.
- Measured brain activity of a given individual may be compared to files from that database of that individual to determine what the individual is attempting to communicate or what type of mental state the individual is in.

## In the Drawings below,

- **FIG. 1 is a block diagram illustrating the system and method according to a first step toward this WSN based direction.**

- **FIG. 2 is block diagram illustrating the neural network inclusive of computer of the FIG.1 stage.**

- **FIGS. 3(a) and 3(b) are exemplary graphs of monitored brain activity of different individuals with, for example, FIG. 3(a) illustrating monitored brain activity of a particular individual who is attempting to communicate the word "NO" and FIG. 3(b) illustrating monitored brain activity of the same individual when that individual is attempting to communicate the word "YES".**

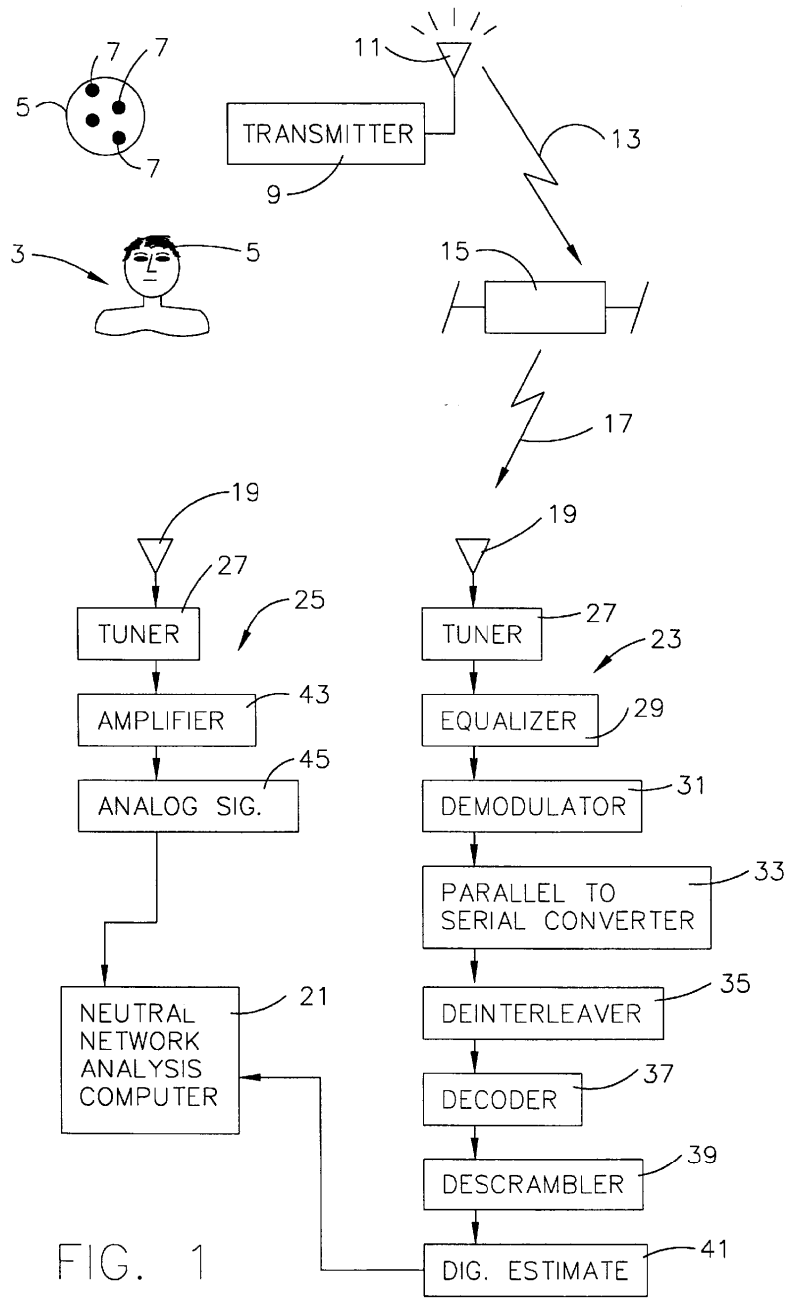

FIG. 1

FIG. 2

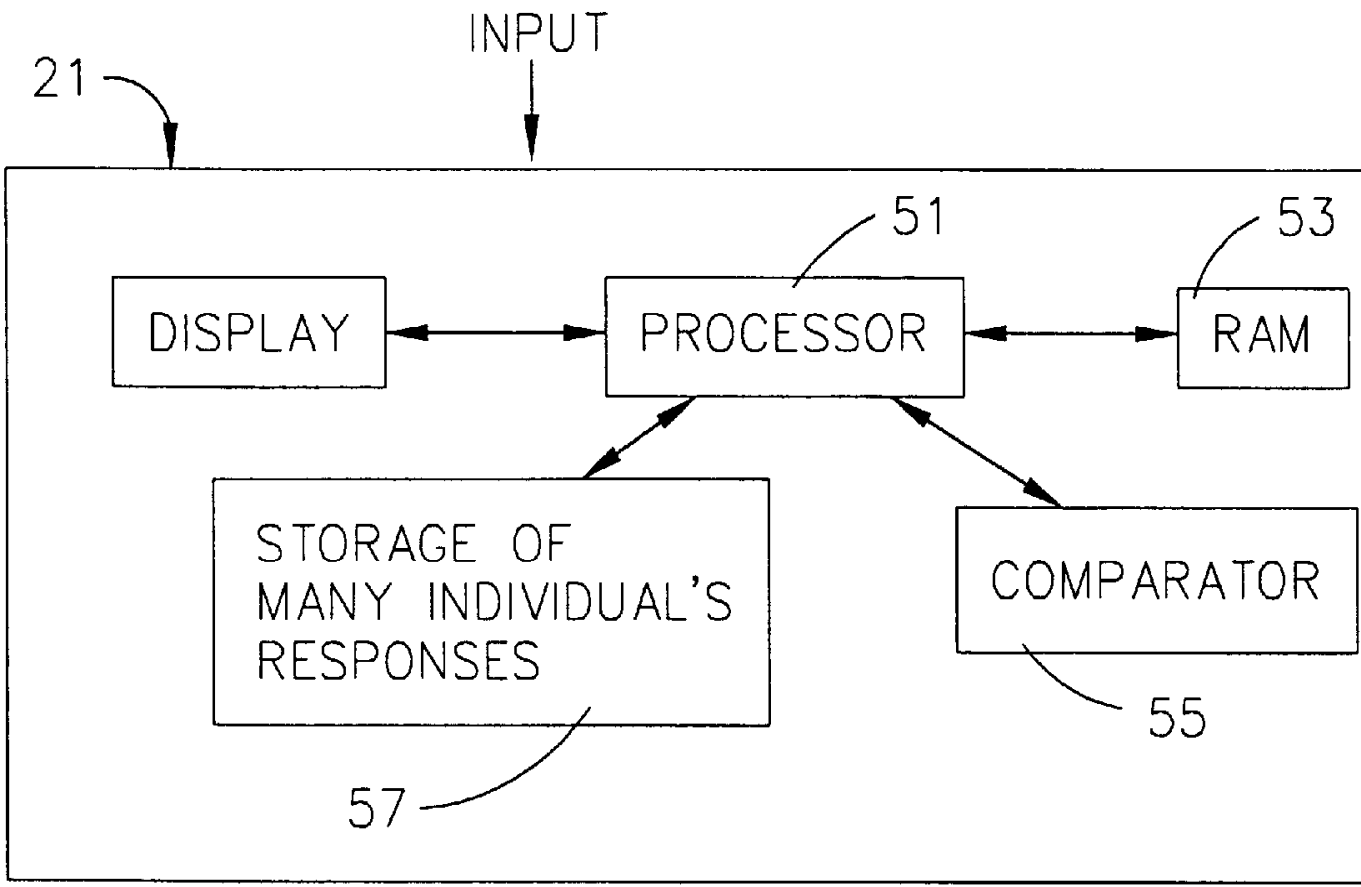

## FIG. 3(a)

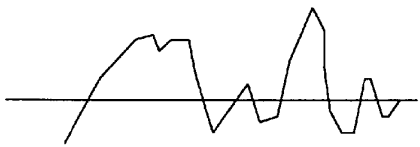

JOE (NO)

## FIG. 3(b)

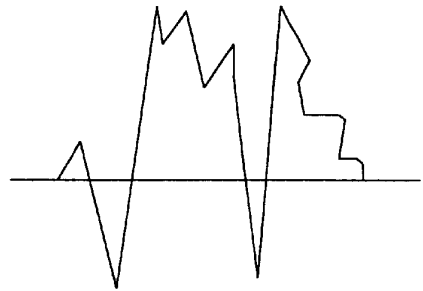

JOE (YES)

It is an object of this WSN based future communication technology direction to enable brain activity of a first human being to be monitored, with the activity being transmitted to a remote location so that individuals and/or a computer at the remote location can determine what the first human being was thinking or intending to communicate. In such a manner, human beings can communicate with one another via monitoring of brain activity, and transmission of the same.

It is another object of this future WSN based communication technology direction to communicate monitored brain activity from one location to another in a wireless manner, such as by IR, RF, or satellite.

In certain terms of this WSN based future communication technology direction, the computer located at the remote location includes a neural network suitably programmed in accordance with known neural network techniques, for the purpose of receiving the monitored brain activity signals, transforming the signals into useful forms, training and testing the neural network to distinguish particular forms and patterns of physiological activity generated in the brain of the monitored individual, and/or comparing the received monitored brain activity information with stored information relating to that individual in order to determine what the individual is attempting to communicate.

*Reference:-*
*Aris Mardirossian: Communication system and method including brain wave analysis and/or use of brain activity US Patent Issued on January 4, 2000*